\documentclass[aps,epsf,subfigure,twocolumn]{revtex4-1}

\usepackage{caption}
\usepackage{graphicx}
\usepackage{epstopdf}
\usepackage{dcolumn}
\usepackage{bm}
\usepackage{amsmath}
\usepackage{color,psfrag,epsfig}
\usepackage{subfigure}
\begin{document}

\newcommand{\bk}{{\bf k}}
\newcommand{\bc}{\begin{center}}
\newcommand{\ec}{\end{center}}
\newcommand{\mub}{{\mu_{\rm B}}}
\newcommand{\sD}{{\scriptscriptstyle D}}
\newcommand{\sF}{{\scriptscriptstyle F}}
\newcommand{\sCF}{{\scriptscriptstyle \mathrm{CF}}}
\newcommand{\sH}{{\scriptscriptstyle H}}
\newcommand{\sAL}{{\scriptscriptstyle \mathrm{AL}}}
\newcommand{\sMT}{{\scriptscriptstyle \mathrm{MT}}}
\newcommand{\sT}{{\scriptscriptstyle T}}
\newcommand{\up}{{\mid \uparrow \rangle}}
\newcommand{\down}{{\mid \downarrow \rangle}}
\newcommand{\upsp}{{\mid \uparrow_s \rangle}}
\newcommand{\downsp}{{\mid \downarrow_s \rangle}}
\newcommand{\upsone}{{\mid \uparrow_{s-1} \rangle}}
\newcommand{\downsone}{{\mid \downarrow_{s-1} \rangle}}
\newcommand{\upt}{{ \langle \uparrow \mid}}
\newcommand{\downt}{{\langle \downarrow \mid}}
\newcommand{\bbar}{{\mid \uparrow, 7/2 \rangle}}
\newcommand{\abar}{{\mid \downarrow, 7/2 \rangle}}
\renewcommand{\a}{{\mid \uparrow, -7/2 \rangle}}
\renewcommand{\b}{{\mid \downarrow, -7/2 \rangle}}
\newcommand{\plus}{{\mid + \rangle}}
\newcommand{\minus}{{\mid - \rangle}}
\newcommand{\psio}{{\mid \psi_o \rangle}}
\newcommand{\psis}{{\mid \psi \rangle}}
\newcommand{\bpsio}{{\langle \psi_o \mid}}
\newcommand{\barpsi}{{\mid \psi' \rangle}}
\newcommand{\barpsio}{{\mid \bar{\psi_o} \rangle}}
\newcommand{\ex}{{\mid \Gamma_2^l \rangle}}
\newcommand{\LH}{{{\rm LiHoF_4}}}
\newcommand{\LHx}{{{\rm LiHo_xY_{1-x}F_4}}}
\newcommand{\de}{{{\delta E}}}
\newcommand{\Ht}{{{H_t}}}

\title{The strain gap in a system of weakly and strongly interacting two-level systems}

\author{A. Churkin$^{1,2}$, I. Gabdank$^{2,\dagger}$, A. L. Burin$^3$, and M. Schechter$^2$} 
\email{alexach3@sce.ac.il}
\email{gabdank@stanford.edu}
\email{aburin@tulane.edu}
\email{smoshe@bgu.ac.il}
\affiliation{$^1$Department of Software Engineering, Sami Shamoon College of Engineering, Beer-Sheva, Israel\\$^2$Department of Physics, Ben-Gurion University of the Negev, Beer Sheva
84105, Israel\\
$^3$Department of Chemistry, Tulane University, New Orleans, LA, 70118, USA}

\altaffiliation{Present address: Department of Genetics, Stanford University School of Medicine, Stanford, California 94305, USA}

\date{\today}

\begin{abstract}
Many disordered lattices exhibit remarkable universality in their low temperature properties, similar to that found in amorphous solids. Recently a two-TLS (two-level system) model was derived based on the microscopic characteristics of disordered lattices.
Within the two-TLS model the quantitative universality of phonon attenuation, and the energy scale of $1-3$ K below which universality is observed, are derived as a consequence of the existence of two types of TLSs, differing by their interaction with the phonon field. In this paper we calculate analytically and numerically the densities of states (DOS) of the weakly and strongly interacting TLSs. We find that the DOS of the former can be well described by a Gaussian function, whereas the DOS of the latter have a power law correlation gap at low energies, with an intriguing dependence of the power on the short distance cutoff of the interaction. Both behaviors are markedly different from the logarithmic gap exhibited by a single species of interacting TLSs.
Our results support the notion that it is the weakly interacting $\tau$-TLSs that dictate the standard low temperature glassy physics. Yet, the power-law DOS we find for the $S$-TLSs enables the prediction of a number of deviations from the universal glassy behavior that can be tested experimentally.
Our results carry through to the analogous system of electronic and nuclear spins, implying that electronic spin flip rate is significantly reduced at temperatures smaller than the magnitude of the hyperfine interaction.
\end{abstract}


\maketitle

\section{Introduction}
\label{Sec:introduction}

A model of two types of two-level systems (TLSs), interacting weakly and strongly with the vacuum field, describes both magnetic insulators having electronic and nuclear spins, and orientational glasses, in which inversion symmetric and inversion asymmetric excitations have a weak and strong interaction with the phonon field, respectively\cite{SS09}.

Such orientational glasses, and similarly amorphous solids, show remarkable universality in their low temperature characteristics; their specific heat increases as $T^a$ with $a \approx 1$, the thermal conductivity increases as $T^b$ with $b  \approx 2$ (both quantities increase as $T^3$ in ordered lattices), and their internal friction is temperature independent\cite{ZP71,HR86,PLT02}. The universality is also quantitative, as all characteristics that are dictated by phonon attenuation suggest a rather similar ratio of $\sim 150$ between phonon mean free path to phonon wavelength, in systems ranging from amorphous solids, to disordered lattices, polymers and porous aerogels\cite{PLT02}. All the above phenomena are present below a temperature of $1-3$K, which is, again, rather  universal.

Much of the universal behavior of disordered systems at low temperatures can be explained by the "Standard Tunneling Model"\cite{AHV72,Phi72,Jac72}. This model introduces phenomenologically tunneling two-level systems (TLSs), which interact weakly with the phonon field and negligibly between each other. TLSs are of fundamental and practical interest: TLSs constitute the dominant dynamic degrees of freedom determining low temperature thermodynamics and kinetics in amorphous solids\cite{DH81,BGH90,RNO96}; TLSs are essential for quantum information and metrology since they restrict the quantum coherence in nano-devices, such as Josephson junction qubits\cite{SLH+04,SSMM05,MCM+05,KPV18,SS19,BJ19,MCL19} and nanomechanical oscilators\cite{Nems08,Nems09,Nems10}; TLSs can be used constructively, e.g. for lasing\cite{RKBO16}. Understanding the TLSs nature can help to reduce their destructive effect, and enhance their constructive usage.

Although the standard tunneling model has been successful in explaining many of the above mentioned characteristics of disordered solids at low temperatures, it left some central questions unanswered, including the explanation of the quantitative universality between different systems, and the origin of the energy scale of $\approx 1-3$K below which universality exists. These questions have been the subject of theoretical scrutiny\cite{KFAA78,KKI83,SC85,AJL88,BNOK98,BGG+92,Par94,LW01,SK94,Kuhn03,PSG07,VL11,Lub18} for the last five decades, yet with limited success.

A novel attempt to resolve these questions was recently made within a "two-TLS" model\cite{SS09}, consisting of two types of TLSs, weakly interacting and strongly interacting with the phonon field. A central property of the two-TLS model is the structure of the low energy density of states (DOS) of the weakly interacting and strongly interacting TLSs. The weakly interacting TLSs have energies smaller than $\approx 10$K, and  are only weakly affected by TLS-TLS correlations. In contrast, the strongly interacting TLSs have typical energies of $\approx 300$K, but are significantly gapped at low energies. As a result, the weakly interacting TLSs dictate the universal phonon attenuation at low temperatures; and the width of their energy distribution, and the consequent gapping of the strongly interacting TLS DOS at the same energy scale, dictates the energy scale of $\approx 1-3$K below which universal properties are observed.

Whereas the two-TLS model explains the above mentioned phenomena as they appear in disordered lattices and amorphous solids alike, for disordered lattices the two-TLS model is microscopically derived\cite{SS09}. Both the form and strengths of the interactions suggested by the two-TLS model and the resulting DOS of the weakly and strongly interacting TLSs are found in excellent agreement with numerical calculations using bare atomic interactions Hamiltonian\cite{GS11,CBS13,CBS14}, the calculated interaction strength being in agreement with experiment\cite{GS11}. Furthermore, for $T < 1-3$ K, where other degrees of freedom are gapped, the two-TLS model can be used to calculate, for disordered lattices, not only the universal phonon attenuation properties, but also properties such as nonlinear acoustic absorption, and phonon attenuation under non-equilibrium created by DC bias \cite{SNB18}. For $T > 3$ K other degrees of freedom, e.g. molecular librations, become significant and dominate the physics of orientational glasses \cite{RS88,GRS88}. However, also in this temperature range it is to be expected that tunneling states affect physical properties, and thus the two-TLS model, and specifically the bimodal structure of the DOS of the bias energies as is calculated below, may be found relevant.

With regard to amorphous solids, the general applicability of the two-TLS model is an open question of current research, which is, however, beyond the scope of this paper. Still, the two-TLS model, and specifically the form of the DOS of the weakly and strongly interacting TLSs found here, have been found relevant in amorphous solids, explaining TLS pure dephasing and non-equilibrium absorption, not
accounted for by the STM \cite{MSSS16,KSB+17}. Furthermore, the existence of two types of TLSs was demonstrated
in amorphous ${\rm Al_2O_3}$ and ${\rm LaAlO_3}$ films \cite{KSG+13}, where the weakly interacting TLSs were attributed to Hydrogen impurities. Additional tests for the applicability of the two-TLS model to amorphous solids were suggested in Ref. \cite{SNB18}.
Perhaps the most direct evidence for the applicability of the two-TLS model in an amorphous solid were recently found in experiments on amorphous silicon. Measuring mechanical and dielectric losses in amorphous silicon films pointed to the existence of two distinct types of TLSs \cite{Hellman21}. Furthermore, recent dielectric loss measurements of amorphous silicon under applied time dependent bias \cite{Liuqi21} cannot be interpreted using the STM, and show good agreement with the two-TLS model, strongly suggesting the existence of weakly and strongly interacting TLSs in amorphous silicon.

The gapping of the strongly interacting TLSs is a manifestation of the Efros-Shklovskii gap for a system of two types of TLSs interacting via a random dipolar-like interaction. The Efros-Shklovskii gap of low energy single particle excitations in glassy systems with long range interactions \cite{ES75} has been a subject of thorough theoretical and experimental studies in various systems such as spin glasses \cite{BKS08}, electron glasses \cite{DLR82,VOP02,BSK+06,AOI11}, proteins \cite{BBFK07}, graphene nanoribbons \cite{DKM+11}, and amorphous solids \cite{AJL88,Bur95,RNO96}. The particular behavior of the gap dictates thermodynamic and transport properties. Thus, and since the two-TLS model suggests that the low energy excitations in magnetic insulators, orientationsl glasses, and possibly also in amorphous solids, are generically given by the weakly and strongly interacting TLSs as described in the Hamiltonian (\ref{H-St}),(\ref{Jcutoff}) below, the rigorous calculation of their DOS is of fundamental interest.

In this paper we derive analytically, and verify numerically, the single particle DOS of the weakly and strongly interacting TLSs as are given by the Hamiltonian in Eqs. (\ref{H-St}),(\ref{Jcutoff}). The TLSs interactions are modeled as having a $1/r^3$ spatial dependence with a random angular dependence, and zero or finite short distance cutoff, accounting for magnetic and elastic interactions respectively.
Whereas for a single species the Efros Shklovskii correlation gap is derived by a self-consistent calculation for the DOS \cite{ES75,BSE80}, the two-TLS structure as presented in the Hamiltonian (\ref{H-St}),(\ref{Jcutoff}) allows analytical derivation within controlled approximations, not invoking self-consistency. Our results are then confirmed using Monte-Carlo simulations. In that, we go beyond the numerical treatment performed in Ref. \cite{SS09}, as in the latter also the numerical calculations assumed the Efross-Shklovskii condition.

The DOS of the weakly interacting TLSs is found to be well described by that of a random field Ising model. The magnitude of the random field is much larger than the interaction between two weakly interacting TLSs, yet much smaller than the typical near neighbor defect-defect interaction in solids. This arises because the effective random field is a result of the interaction of the weakly interacting TLSs with the strongly interacting TLSs, thus attaining an intermediate energy scale. The DOS of the strongly interacting TLSs has a {\it power law} dependence on energy at small energies, with a power that depends on the value of the short distance cutoff of the elastic interaction. This dependence of the DOS on the form of the interaction at short distances is quite remarkable, as the physics of the correlation gap at low energies is that of long distances. We show that the product of the $r^{(-3)}$ interaction constant at large distances and the DOS of the weakly interacting TLSs at low energies determines the gap behavior. This product depends on the short distance cutoff, because with enhanced cutoff the distribution of energies of the weakly interacting TLSs becomes narrower, and their DOS at low energies becomes larger.

The paper is organized as follows: In Sec.~\ref{Sec:Analytical} we present the model and the analytical derivation of the DOS of the weakly interacting and strongly interacting TLSs. Our numerical results are presented in Sec.~\ref{Sec:Numerical}. We then conclude with a discussion of our results and their consequences in Sec.~\ref{Sec:Discussion}. Some details of the analytical derivation of the DOS are deferred to App.~\ref{Sec:tauDOS} (weakly interacting TLSs) and App.~\ref{Sec:SDOS} (strongly interacting TLSs). The numerical calculation of the power law in the DOS of the $S$-TLSs is detailed in App.~\ref{Appnumerical}, and the effect of finite size is considered in App.~\ref{Appfinitesize}.

\section{Analytical theory}
\label{Sec:Analytical}

The two-TLS model assumes a bi-modality of the interaction of TLSs with the mediating field. We refer below to two types of structural TLSs having weak and strong interactions with the phonon field, but a complete analogy exists with nuclear and electronic spins interacting with the electromagnetic vacuum. This form of the bimodal interaction, which, for the disordered lattices, is rigorously derived\cite{SS09} (see also Refs. \cite{GS11,CBS13}), leads to acoustic phonon mediated interaction between tunneling TLSs in disordered systems that is described by the Hamiltonian\cite{SS08,SS09}

\begin{multline}
H_{S \tau} \;=\; -\sum_{i \neq j} \left[\frac{1}{2}J_{ij}^{SS} S_i^z S_j^z +
J_{ij}^{S\tau} S_i^z \tau_j^z + \frac{1}{2}J_{ij}^{\tau \tau} \tau_i^z
\tau_j^z \right] \, ,
 \label{H-St}
\end{multline}
where
\begin{equation}
J_{ij}^{ab}=c_{ij}^{ab} \frac{J_o^{ab}}{r_{ij}^3 + \tilde{a}^3} \, .
\label{Jcutoff}
\end{equation}
$a,b=S$ or $\tau$ stand for the strongly interacting and weakly interacting TLSs, and are denoted by pseudo Ising spins with $S^z=\pm 1, \tau^z=\pm 1$. $c_{ij}^{ab}$ is chosen randomly from a Gaussian distribution of width unity. The factor of $1/2$ in the first and third terms of the Hamiltonian (\ref{H-St}) accounts for the double summation.
$J_o^{ab}$ denote the interaction energy scales at nearest neighbor distance, and are related by $J_o^{\tau \tau} = g J_o^{S \tau} = g^2 J_o$\cite{SS09}, where we define $J_o \equiv J_o^{SS}$ and the dimensionless parameter, considered small throughout this paper, $g \ll1$. This separation of energy scales, which distinguishes between the weakly interacting $\tau$-TLSs and the strongly interacting $S$-TLSs, is central to our analysis. In orientational glasses $J_o$ is typically of the order of the Debye energy, and $g \approx 0.01-0.03$\cite{SS09,GS11,CBS13}.
Both the elastic interaction in glasses and the magnetic dipolar hyperfine interaction decay at large distances as $1/r^3$, but have a different spatial dependence at distances of the order of the interatomic distance. We therefore introduce a short distance cutoff $\tilde{a}$ through Eq.(\ref{Jcutoff}). We expect other forms of the cutoff, that eliminate the divergence of the interaction at small distances, to give similar results. Here and throughout the paper we assume that the $S$-TLSs and $\tau$-TLSs are randomly placed in a lattice with concentration (density per site) $\rho$, and distances will be given in units of the lattice spacing $a_o$.

We emphasize that in this paper we are concerned only with the bias energies $E$ (usually denoted $\Delta$) between the two states of each of the TLSs, and not with the tunneling amplitudes $\Delta_o$ between these states, which can be approximately ignored for the dipole gap analysis. As such, the DOS calculated for both the weakly interacting TLSs and the strongly interacting TLSs is that of the bias energies. The full energy of a TLS is given by $\sqrt{E^2+\Delta_o^2}$. We then postulate the distribution functions in the standard form

\begin{equation}
P^{(\tau,S)}(E,\Delta_o)=P^{(\tau,S)}_o(E)/\Delta_o \, ,
\end{equation}
following the ansatz of the standard tunneling model for the independence of the bias energies and the tunneling amplitudes, but allowing an energy dependence of the DOS of the bias energies, which we calculate below. We note that while the typical values of the tunneling amplitudes are limited below a few Kelvin, their range
(i.e. $\Delta_o^{min}$ and $\Delta_o^{max}$) could differ between the $\tau$-TLSs and S-TLSs. However, the $1/\Delta_o$ dependence of the distribution function rests on the approximate homogeneity of the tunnel barriers, which is a generic consideration similarly applicable to both classes of tunneling states. For a further discussion of the approximation neglecting $\Delta_o$ in the calculation of the TLS DOS see Ref. \cite{CMBS21}.

The Hamiltonian we consider [in Eq. (\ref{H-St})] is therefore classical, and the bias energies are given by the single particle excitation (spin flip) energies for the $S$-TLSs and for the $\tau$-TLSs as

\begin{equation}
E_S^j \equiv 2\sum_{i(\neq j)}[J_{ij}^{SS} S_i^z S_j^z +
J_{ij}^{S\tau} S_j^z \tau_i^z]
\label{Senergy}
\end{equation}
and
\begin{equation}
E_{\tau}^j \equiv 2\sum_{i(\neq j)}[J_{ij}^{S\tau} S_i^z \tau_j^z + J_{ij}^{\tau \tau} \tau_i^z
\tau_j^z] .
\label{tauenergy}
\end{equation}
The factor of $2$ is a result of the spin changing by $2$ upon flipping.
We then calculate for the $\tau$-TLSs and for the S-TLSs the functional dependence of the single particle DOS at zero temperature, $n_{\tau}(E)$, $n_S(E)$. The uncorrelated DOS for the $S$-TLSs and $\tau$-TLSs is given by a Gaussian distribution with typical energy and peak values dictated by the dominating interaction ($S-S$ for the $S$-TLSs and $S-\tau$ for the $\tau$ TLSs). However, correlations in between the $S$-TLSs, between the $S$-TLSs and the $\tau$ TLSs, and in between the $\tau$-TLSs result in a depression in $n_S(E)$ and in $n_{\tau}(E)$ at low energies. This depression is governed by the Efros-Shklovskii stability criterion\cite{ES75}, which for the $S-S$ correlations reads:

\begin{equation}
E_S^i + E_S^j - 4J_{ij}^{SS} S_i^z S_j^z > 0 \,
\label{ESSS}
\end{equation}
for any two $S$-TLSs in the system, while for the $S-\tau$ correlations one requires

\begin{equation}
E_S^i + E_{\tau}^j - 4J_{ij}^{S\tau} S_i^z \tau_j^z > 0 \,
\label{ESStau}
\end{equation}
for any $S$ and $\tau$ TLSs, and for the $\tau-\tau$ correlations one has

\begin{equation}
E_{\tau}^i + E_{\tau}^j - 4J_{ij}^{\tau\tau} \tau_i^z \tau_j^z> 0 \, .
\label{EStautau}
\end{equation}

These conditions are a manifestation of the requirement that the ground state must be stable to flips of any pair of spins. Due to the many-body nature of spin-spin interactions the energy to flip such a pair can be lower than the sum of the single TLS flip energies. The difference is represented by four times the intra pair interaction term (a factor of $2$ for each single spin flip as in Eqs. (\ref{Senergy}),(\ref{tauenergy}), and another factor of $2$ comes from the fact that this interaction energy is counted once for each single flip, but not at all for the double flip). For a single species of TLSs interacting via the dipolar interaction, e.g. for a system described by the Hamiltonian given by the first term in Eq.(\ref{H-St}) with $\tilde{a}=0$, correlations lead to logarithmic depression of the DOS, which for low energies $E$ is proportional to $1/\log{(E_o/E)}$\cite{BSE80,Bur95}. The same of course would be true for the $\tau$-TLSs in the absence of the $S$-TLSs. However, the presence of two types of TLSs, with significant difference in their coupling as is given in Eq.(\ref{H-St}), changes things remarkably.

\subsection{DOS of the weakly interacting TLSs}

Let us consider first $n_{\tau}(E)$. As is shown below, $S$-TLS excitations are scarce at low energies, and thus, for the consideration of the $\tau$-TLSs, to first approximation, and up to an overall constant (resulting from the $S-S$ interactions),
the Hamiltonian in Eq.(\ref{H-St}) reduces at low temperatures to the effective Hamiltonian

\begin{multline}
H_{\tau}^{RF} \;=\; -\sum_{i \neq j} \left[J_{ij}^{S\tau} \langle S_i^z \rangle \tau_j^z + \frac{1}{2} J_{ij}^{\tau \tau} \tau_i^z
\tau_j^z \right] \\ \equiv -\frac{1}{2} \sum_{i \neq j} J_{ij}^{\tau \tau} \tau_i^z \tau_j^z + \sum_j h_j \tau_j^z \, ,
\label{tauRFIM}
\end{multline}
which is equivalent to the random field Ising model with the random field $h_j = -\sum_{i(\neq j)} J_{ij}^{S\tau} \langle S_i^z \rangle$ being $\approx 1/g$ times larger than the $\tau-\tau$ interaction.
The large random field leads to a substantial reduction of the effect of the $\tau-\tau$ correlations.  The effect of the correlations with the $S$-TLSs [present in the full Hamiltonian (\ref{H-St})] is even much weaker because of the smallness of $n_S(E)$ at low energies, as is derived below.
In Ref.\cite{SS09} it was shown that $n_{\tau}(E)$ dips because of correlations only at energies smaller than $g^2 J_o$, and that the relative decrease in DOS is small, proportional to $g$. Apart from this decrease we show in App.\ref{Sec:tauDOS} that for intermediate TLS spatial concentrations $\rho$, i.e. $[z(1+\tilde{a}^3)]^{-1} < \rho \lesssim 1$, $n_{\tau}(E)$ is well approximated by a Gaussian distribution\cite{Lorenote}.
Here $z$ is the coordination number of the lattice (or the underlying lattice in the amorphous state), and all distanced are in units of interatomic spacing $a_o$ taken to be equal $1$. The left inequality ensures that the few largest contributions (coming from $\tau-S$ interactions) to any given $\tau$-TLS are of similar magnitude. The second inequality assures strong enough disorder. The width of the distribution, $E_{\tau}^{typ}$, is dictated by the Hamiltonian (\ref{tauRFIM}). In the limit of $\tilde{a} \gg a_o$ we find

\begin{equation}
n_{\tau}(E) = \frac{2 \rho}{\sqrt{2\pi}E_{\tau}^{typ}}e^{\frac{E^2}{2(E_{\tau}^{typ})^2}}
\end{equation}
with
\begin{equation}
E_{\tau}^{typ}=\sqrt{\frac{16\pi}{3} \frac{\rho}{\tilde{a}^3}} g J_o \, .
\label{E-typ}
\end{equation}
The analysis of the distribution in detail is given in App.\ref{Sec:tauDOS}. Both the Gaussian functional dependence of $n_{\tau}(E)$ and the analytical results for the widths of the distribution as function of $\tilde{a}$ are verified numerically, see Sec. \ref{Sec:Numerical}.

\subsection{DOS of the strongly interacting TLSs}

We now turn to the $S$-TLSs. At energies much larger than $gJ_o$, i.e. energies larger than the maximum energy of a $\tau$-TLS excitation, we expect the $S$-TLS DOS to have the same functional behavior, i.e. logarithmic depression, as in the absence of the $\tau$-TLSs. This is confirmed numerically in Sec. \ref{Sec:Numerical}. However, at low energies, where $S$-TLSs are scarce and $\tau$-TLSs are abundant, $S-\tau$ correlations as manifested in Eq.(\ref{ESStau}) dominate the functional dependence of $n_S(E)$. We note also that it is the low energy regime which is of most interest to us, as in the related regime of low temperatures TLSs determine the physical properties of glasses.
We thus calculate the $S$-TLS DOS resulting from Eq.(\ref{ESStau}) analytically for energies $E \ll gJ_o$.

\begin{figure*}
\subfigure[]{
  \includegraphics[width=8.6cm]{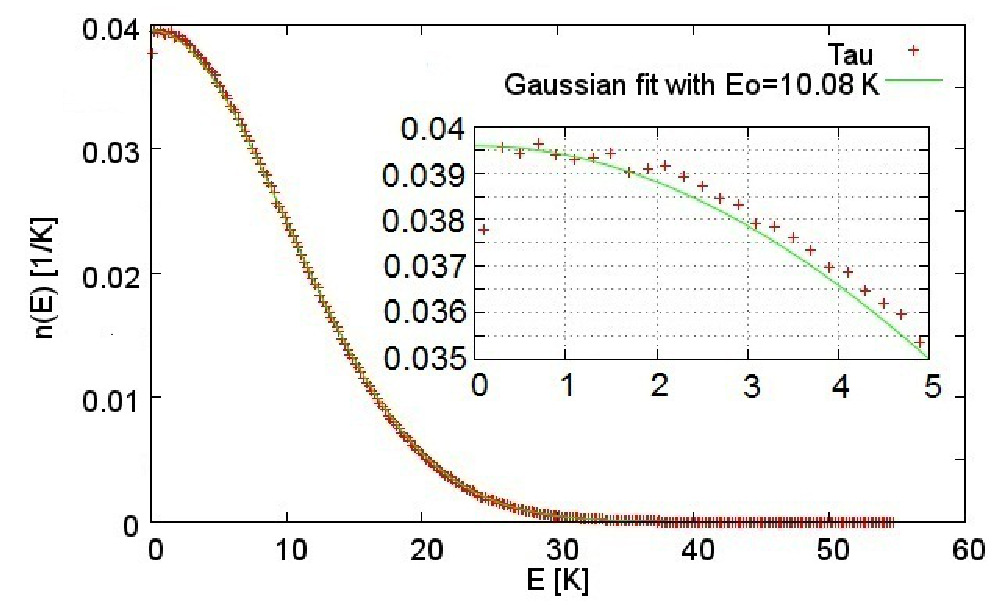}
   \label{fig:Fig1subfig1}
   }
 \subfigure[]{
  \includegraphics[width=8.6cm]{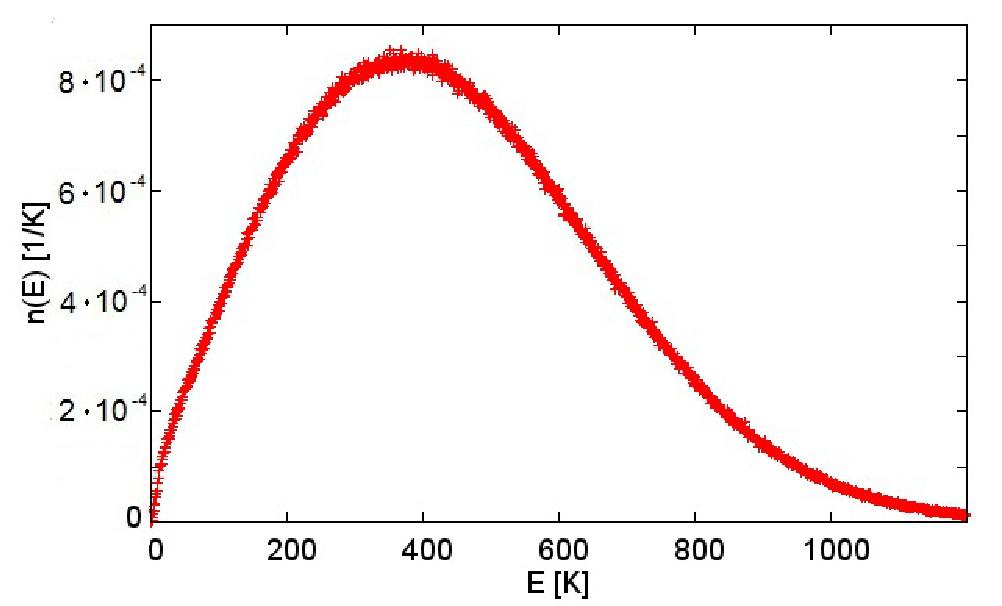}
   \label{fig:Fig1subfig2}
   }
 \subfigure[]{
  \includegraphics[width=8.6cm]{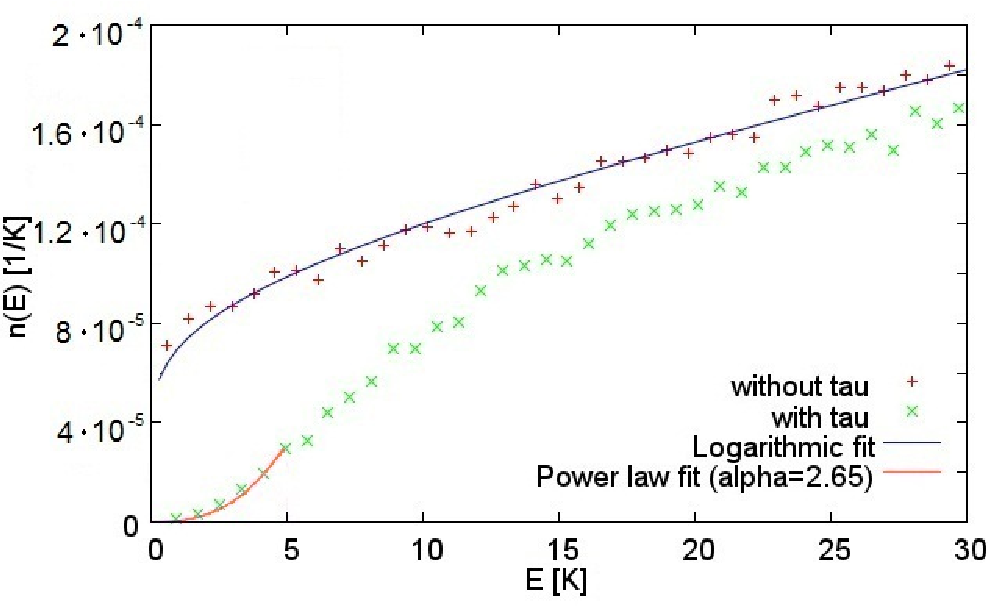}
   \label{fig:Fig1subfig3}
   }
 \subfigure[]{
  \includegraphics[width=8.6cm]{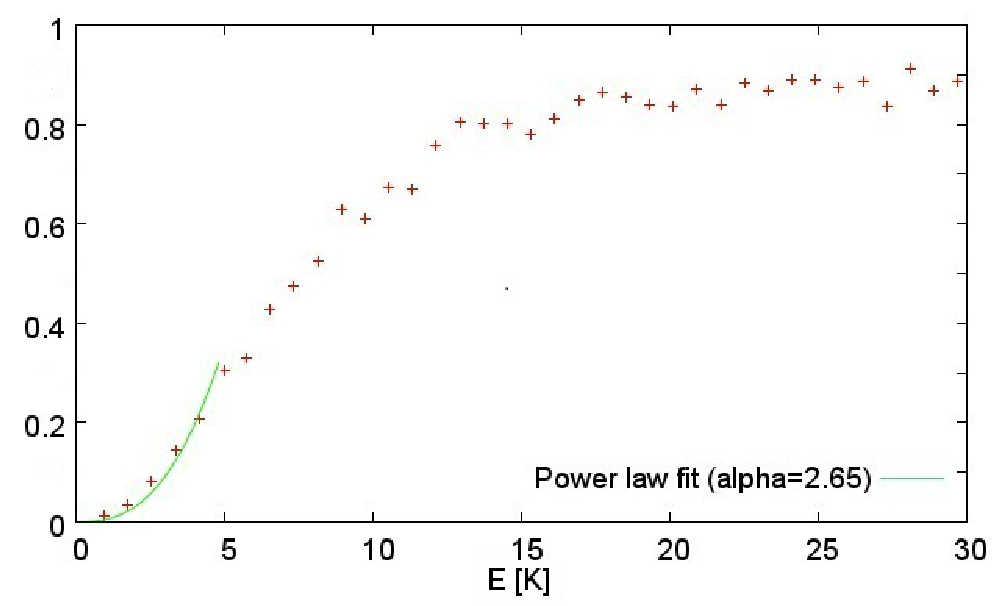}
   \label{fig:Fig1subfig4}
   }
\caption{ Single particle DOS for cube of volume $12\times12\times12$ with $\rho=0.5$ and $\tilde{a}=2$. [Here and in all subsequent figures DOS is plotted per unit cell (i.e. per site)](a) $\tau$-TLSs DOS with zero free parameters Gaussian fit, using the standard deviation calculated in App. \ref{Sec:tauDOS}. Inset shows the small shift of states from the lowest energies upwards. (b) $S$-TLSs DOS.  (c)  Zoom into low energies of the $S$-TLSs DOS in the absence and presence of $\tau$-TLSs. The shift from log to power law behavior as a result of the $S-\tau$ interactions is presented. Solid lines denote a fit to the function $A/Log(B/E)$ [$A=250, B=0.00038$, upper curve], and to the analytical result for the $S$-TLS DOS at low energies, [Eq.(\ref{PE6}) with $\alpha=2.65$, lower curve]. (d) The ratio between the $S$-TLSs DOS with $\tau$ TLSs and
the $S$-TLSs DOS  in the absence of the $\tau$ TLSs, zoomed into low energies.}
\label{fig:Fig1}
\end{figure*}

We define $P(E)$ by

\begin{equation}
n_S(E) = P(E) \bar{n}_S(E)
\label{PEdef}
\end{equation}
where $\bar{n}_S(E)$ is the $S$-TLS DOS neglecting $S-\tau$ correlations, and calculate explicitly $P(E)$.
We use the Efros-Shklovskii criterion of stability

\begin{equation}
E_{\tau}+E_S-\frac{4\eta gJ_o}{R^3+\tilde{a}^3} > 0
\label{E_S}
\end{equation}
which, for each $S$-TLS, should be fulfilled for all $\tau$-TLSs. Thus, for a given $S$-TLS, the reduction factor in its probability to be at an energy $E$ coming from fulfilling Eq.(\ref{E_S}) for all $\tau$-TLSs is given by

\begin{multline}
P(E)=\prod_{d{\bf R}} \biggl[ 1-d{\bf R}\int_{\frac{E\tilde{a}^3}{4gJ_o}}^{\infty} d\eta \frac{1}{\sqrt{2\pi}} e^{-\eta^2/2} \\ \cdot
\int_0^{4\eta gJ_o/\tilde{a}^3} dE_{\tau} n_{\tau}(E) \Theta\left(-E-E_{\tau}+\frac{4\eta gJ_o}{R^3+\tilde{a}^3}\right) \biggr]  \, .
\label{PE1}
\end{multline}

Since we are interested in low energies we approximate $n_{\tau}(E)$ by a constant given by the $\tau$-TLS DOS at zero energy $n_{\tau}(0)$. Substituting $\zeta \equiv R^3$ we obtain

\begin{multline}
P(E)=exp\biggl[-\frac{4\pi}{3}\int_0^{\infty}d\zeta\int_{\frac{E\tilde{a}^3}{4gJ_o}}^{\infty} d\eta \frac{1}{\sqrt{2\pi}} e^{-\eta^2/2} \\
\cdot \int_0^{4\eta gJ_o/\tilde{a}^3} dE_{\tau} n_{\tau}(0) \Theta\left(-E-E_{\tau}+\frac{4\eta gJ_o}{\zeta+\tilde{a}^3}\right)\biggr] \, .
\label{PE2}
\end{multline}

The detailed evaluation of $P(E)$ is given in App.\ref{Sec:SDOS}. We find that for all values of $\tilde{a}$ the DOS of the $S$-TLSs has a power law dependence on energy at low energies

\begin{equation}
n_S(E) = \bar{n}_S \left( \frac{cE}{gJ_o} \right)^{\alpha} \, ,
\label{PE6}
\end{equation}
where $\alpha$ is given by

\begin{equation}
\alpha=\frac{8\sqrt{2\pi} n_{\tau}(0) gJ_o}{3} \, ,
\label{alphaanalytical}
\end{equation}
and $c$ is a constant of order unity, a result of the ability to calculate the exponent in Eq.~(\ref{PE2}) only within logarithmic accuracy.
The dependence of $\alpha$ on the interaction cutoff $\tilde{a}$ is given through $n_{\tau}(0)$. A general derivation of $n_{\tau}(0)$, along with the values of $\alpha$ for various cutoffs $\tilde{a}$ and concentrations $\rho$, is given in App. \ref{Sec:tauDOS}, see also Table \ref{tab:Table2}.

For large short distance cutoffs, $\tilde{a} \gg a_o$ we find

\begin{equation}
n_{\tau}^{\tilde{a}}(0) = \frac{\sqrt{3/2}\sqrt{\rho \tilde{a}^3}}{4\pi gJ_o} \, .
\label{ntauzero}
\end{equation}
We assume here, for simplicity, $\rho_S=\rho_{\tau}=\rho$, but generalization to $\rho_S \neq \rho_{\tau}$ is straightforward, see App. \ref{Sec:tauDOS}. We can then write the $S$-TLS exponent as

\begin{equation}
\alpha=\frac{2 \sqrt{\rho \tilde{a}^3}}{\sqrt{3\pi}} \, .
\end{equation}
With decreasing $\tilde{a}$, $\alpha$ decreases monotonically, reaching the value of $\alpha=0.92 \sqrt{\rho}$ for $\tilde{a}=0$.

\section{Numerical calculations}
\label{Sec:Numerical}

\begin{figure*}[!tbh]
\centering
\subfigure[]{
  \includegraphics[width=8.6cm]{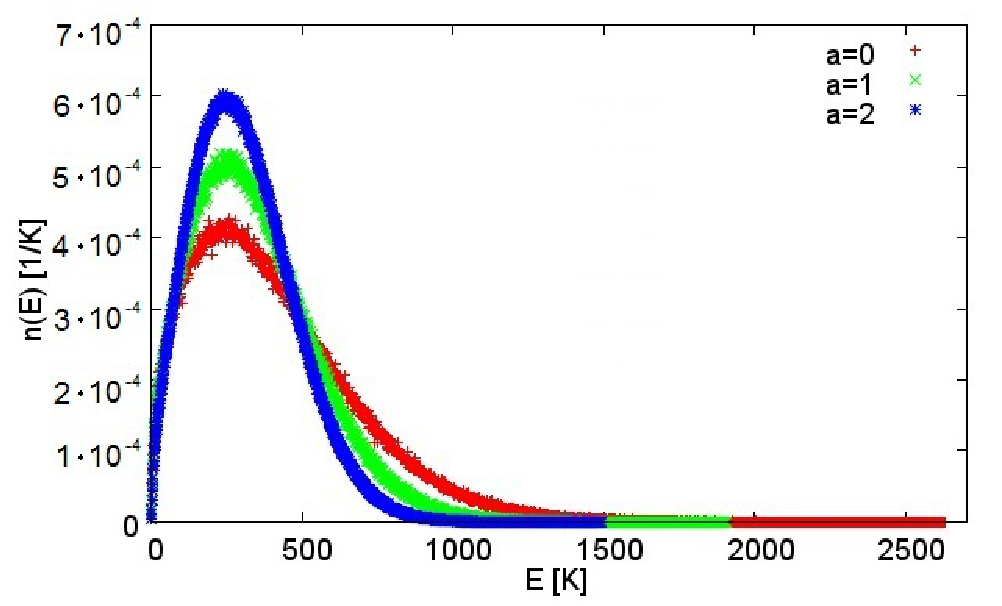}
   \label{fig:Fig2subfig1}
   }
 \subfigure[]{
  \includegraphics[width=8.6cm]{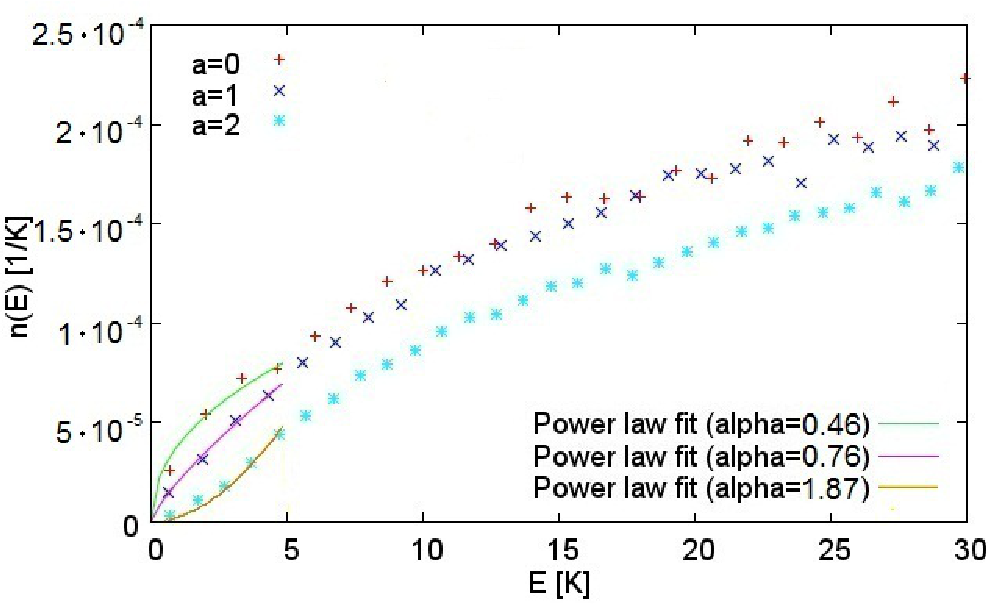}
   \label{fig:Fig2subfig2}
   }
\caption{$S$-TLSs DOS for a cube of volume $14\times14\times14$ with $\rho=0.25$ and different values of interaction cutoff, $\tilde{a}=$0, 1, 2. $g=1/30$ for all cutoffs. $J_o$ is chosen for each value of $\tilde{a}$ such that the peak of the distribution is cutoff independent (see text). (a) $S$-TLSs DOS - entire plot. (b)  Zoom into low energies: 0K - 30K. The increase of the power $\alpha$ with increasing values of interaction cutoff is clearly observed (fits done with $\alpha$ values calculated analytically, as are given in Table~\ref{tab:Table2}).}
\label{fig:Fig2}
\end{figure*}

Our analytical derivation of the DOS relies on the Efros Shklovskii criterion for the low energy DOS of the $S$-TLSs, and on the assumption of a Gaussian distribution for all but the low energy excitations of the $\tau$-TLSs. In order to check our analytical results and verify the validity of the above assumptions we perform a numerical calculation of the DOS of both the $\tau$-TLSs and the $S$-TLSs starting from the Hamiltonian in Eqs.(\ref{H-St}),(\ref{Jcutoff}).

\subsection{DOS of $S$ and $\tau$ TLSs}

Calculations are performed on cubic lattices of size $L$, with $L=6-14$ and periodic boundary conditions, where the interaction between each pair of TLSs is calculated according to their shortest separation in the extended lattice. $S$ and $\tau$ TLSs are placed randomly in the lattice with concentration $\rho$ where the cases of $\rho=0.25$ and $\rho=0.5$ are analyzed. In accordance with the model in Ref. \cite{SS09} we take each occupied site to contain both an $S$ and a $\tau$ TLS, and avoid on site interactions, but taking the positions of the $S$ and $\tau$ TLSs to be uncorrelated produces similar results. Using Monte Carlo simulation with the Hamiltonian (\ref{H-St}),(\ref{Jcutoff}) we lower the free energy of the system at temperatures decreasing from $300$K to $0.01$K. As in realistic systems, our simulations reach very low energy states at the lowest temperature, yet the system does not equilibrate. Such low energy states produce the correct Efros Shklovskii gap once they are stable to single and double spin flips\cite{BSE80}. Both conditions are explicitly checked, and are very well satisfied in our simulations at $T=0.01$K. Once we reach the final state of the simulation at a given size, dilution, and interaction cutoff, we measure the excitation energy of each $S$ and $\tau$ TLS.

In Fig.~\ref{fig:Fig1}(a),(b) we plot the single particle DOS of the $\tau$-TLSs and of the $S$-TLSs, calculated for a cube of volume $12\times12\times12$ with impurity concentration $\rho=0.5$ and short distance cutoff $\tilde{a}=2$, with the parameters $J_o=300$K and $g=1/30$. The typical energy scale for $S$ excitations is indeed $J_o$, a result of the interactions between different $S$-TLSs. At lower energies $n_S(E)$ is logarithmically reduced. The typical energy scale of the $\tau$-TLSs is of order $gJ_o$, a result of their interactions with $S$-TLSs.
In Fig.\ref{fig:Fig1}(a) we compare our data to the analytical results presented in Sec.~\ref{Sec:Analytical}, i.e. to a Gaussian with standard deviation of $1.008gJ_o$, see App.\ref{Sec:tauDOS}, Table~\ref{tab:Table1}. We obtain an excellent fit, with no free parameters,
except at very low energies, smaller than $g^2 J_o$, where the DOS is slightly diminished (by relative magnitude $\approx g$) because of $\tau-\tau$ correlations\cite{SS09}. Furthermore, the peak value of $n_{\tau}(E)$ is $\approx 1/g$ times larger than the peak value of $n_S(E)$, as expected\cite{SS09}. This validates the fact that $\tau$-TLSs can be described by the random field Ising model, see Eq.~(\ref{tauRFIM}), with the exception of a small correction at low energies.

The same energy scale, $gJ_o$, which marks the onset of $\tau$-TLSs, marks also the sharp decay in $n_S(E)$.
In Fig.\ref{fig:Fig1}(c) we zoom into low energies. We see that at the energy where the $\tau$ TLSs appear ($\approx 10$K) the reduction of the $S$ TLS DOS changes its functional form to a power law,
as is demonstrated by a fit of the low energy data to Eq.(\ref{PE6}), with $\alpha=2.65$ (see Table~\ref{tab:Table2}).
The fact that this power law gap is a result of the $S-\tau$ correlations is further demonstrated by the calculation of the $S$-TLS DOS in the absence of the $\tau$ TLSs [i.e. taking $J_o^{S \tau} = J_o^{\tau \tau} =0$ in the Hamiltonian (\ref{H-St})]. This graph is shown for comparison in Fig.\ref{fig:Fig1}(c). Indeed, in the absence of $\tau$-TLSs the logarithmic gap continues to low energies. In Fig.\ref{fig:Fig1}(d) we plot the ratio between the DOS of the $S$-TLSs in the absence and in the presence of the $\tau$-TLSs, singling out the effect of the $S-\tau$ correlations on the $S$-TLS DOS.

\subsection{Dependence of the DOS on the short distance cutoff}

In Fig.~\ref{fig:Fig2} we plot $n_S(E)$ for a cube of volume $14\times14\times14$ for different values of interaction cutoffs $\tilde{a}$. For each value of the cutoff, $J_o$ is chosen in the way that the position of the peak of $n_S(E)$ is cutoff independent, and we keep $g=1/30$ independent of the cutoff.
At low energies we indeed find a deepening of the gap with the power $\alpha$ increasing with increasing cutoff $\tilde{a}$, with an excellent fit with the values of $\alpha$ obtained analytically in Sec. \ref{Sec:Analytical}. We emphasize that the value of the power $\alpha$ dictating $n_S(E)$ at low energies does not depend on our choice of $J_o$, as can be inferred from Eq.~(\ref{E-typ}) and Eq.~(\ref{alphaanalytical}) above (noting that $n_{\tau}(0) \propto 1/J_o$).

\begin{figure}[h!]
\centering
\includegraphics[width=8.6cm]{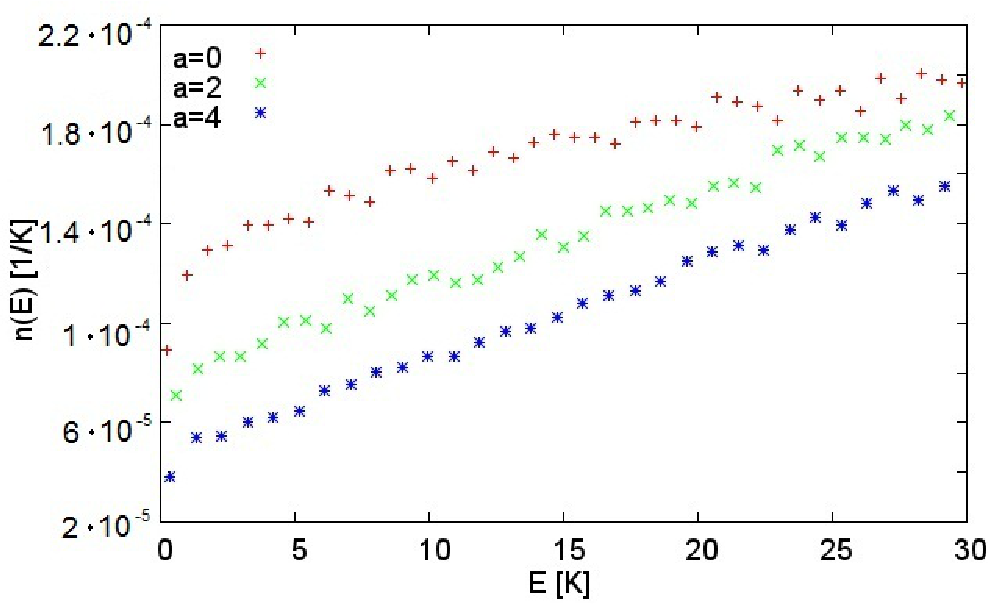}
\caption{$S$-TLSs DOS plots in the absence of the $\tau$ TLSs for cube of volume $12\times12\times12$ with $\rho=0.5$ and $\tilde{a}=0,2,4$. Enhanced cutoffs diminish the low energy DOS, but do not change its functional dependence.}
\label{fig:notau}
\end{figure}

The power $\alpha$ can also be extracted from our numerical results by analyzing the integral plots of the $S$-TLS  DOS. This is done in App.\ref{Appnumerical} for various cutoff parameters, with $\rho=0.25$ and $\rho=0.5$.
Despite finite size effects (see App.\ref{Appfinitesize}) our numerical results for the value of $\alpha$ are in reasonable agreement with the analytical results both in absolute value (see Table \ref{tab:Table2}), and in the functional dependence of $\alpha \propto \sqrt{\rho \tilde{a}^3}$ predicted analytically for $\tilde{a} \gg a_o$.

The specific parameters chosen for $J_o$ and $g$ are in accordance with plausible values for amorphous solids and disordered lattices that express the low energy universal characteristics \cite{ZP71,HR86,PLT02}, and specifically with calculated values for KBr:CN \cite{GS11,CBS13}. Our results show that with these parameters $N_S(E) = g^2 N_{\tau}(E)$ at $E \approx 3$K for cutoffs $\tilde{a}$ of order unity, in agreement with experiment.

The short distance cutoff affects also a system consisting of a single species of TLSs, i.e. having only the first term in the Hamiltonian~(\ref{H-St}). However, in this case the short distance cutoff only changes the magnitude of the DOS at low energies, but not their functional form.  This can be seen in Fig.\ref{fig:notau}, where $N_S(E)$ in the absence of $\tau$ TLSs is plotted for $\tilde{a}=0,2,4$.

\section{Discussion}
\label{Sec:Discussion}

We have derived analytically and numerically the DOS of the bias energies of the weakly ($\tau$) and strongly (S) interacting TLSs within the two-TLS model. We find that the $\tau$-TLSs are confined to energies smaller than $J_o^{S \tau}$ $(\approx 10$ K), whereas the $S$-TLSs are spread to much larger energies, and have a power law gap at low energies.
At energies smaller than $\approx 3$ K, where the $S$-TLSs are scarce, the DOS of the $\tau$ TLSs is, to a very good approximation, constant, and in a wide regime of energies $J_o^{\tau \tau} < E < J_o^{S \tau}$ the $\tau$ TLSs are practically non interacting.
As such, the characteristics of the $\tau$-TLSs at low energies are equivalent to those introduced phenomenologically in the Standard Tunneling Model\cite{AHV72,Phi72}. However, within the two-TLS model [Eq.~(\ref{H-St})] these characteristics are derived, and the relation between the density of states and the coupling to the phonon field is obtained in terms of the small parameter $g$ (Ref. \cite{SS09}). This allows the derivation of the magnitude $C_o \approx 10^{-3}$ of the tunneling strength \cite{PLT02} and its universality, and the energy scale of $gJ_o\approx 3$ K dictating the temperature below which universality is observed \cite{SS09}.

Our results for the DOS of the $S$-TLSs can be used to facilitate predictions for the behavior of various properties of orientational glasses beyond the phenomena of the low temperature universality, both below and above the temperature of $3$ K. Below $3$ K the number of thermal $S$ TLSs is small. Yet, their strong interaction with the phonon field can lead to their dominance of properties with stronger than quadratic dependence on the interaction \cite{MSSS16,KSB+17,SNB18}. At temperature larger than $3$ K it is expected that TLS contribution will be dominated by the $S$-TLSs, resulting in added contribution to that of other prevalent excitations (e.g. librations) at these energy scales, both directly by $S$-TLSs, and through their interactions with e.g. librational modes. Our results here could therefore be useful to the discussion of phenomena such as the plateau in the thermal conductivity at $3-10$ K and the boson peak, as the introduced $S$-TLSs enhance phonon attenuation and the specific heat in the relevant temperature regime.

The two-TLS model was rigorously derived\cite{SS09} and thoroughly validated\cite{GS11,CBS14,CBS13} for the disordered lattices, which constitute a significant subclass of systems showing universality. At the same time, it provides an explanation for crucial aspects of the low temperature universality as are exhibited in both disordered and amorphous solids. Given the equivalence of the phenomenon as observed experimentally in all systems showing the low temperature universality\cite{YKMP86,LVP+98}, we believe it is plausible that the Hamiltonian in Eqs. (\ref{H-St}),(\ref{Jcutoff}) below describes the DOS of the low energy excitations in all amorphous solids.
A direct way to check this possibility would be by detecting the existence of the strongly interacting TLSs, and comparing their DOS to our results here. Specifically, far from equilibrium one can suppress the gap of the S-TLSs at low energies, and consequently they will dominate in acoustic, and even in dielectric response. Indeed, in a recent experiment \cite{Liuqi21} fast bias sweeping of TLSs in amorphous silicon resulted in enhanced S-TLS DOS at low energies, and their dominant contribution  to the dielectric response. Repeating such measurements in other amorphous materials, for dielectric and for acoustic responses, would allow the detection and characterization of weakly and strongly interacting TLSs in various amorphous solids. Other possible experimental verification of the applicability of the two-TLS model to amorphous systems, and specifically the detection of the strongly interacting TLSs, could be done at high energies using e.g. TeraHertz absorption and spectroscopy experiments\cite{PHC+13}, and at low energies using the recently acquired ability to study single TLSs via their interaction with phase qubits and with strain\cite{LMC+10,GPL+12,Bilmes21}, which allows distinction between weakly and strongly interacting TLSs.
Such verification of the applicability of our results to amorphous solids
could give strong support to the general applicability of the two-TLS model in describing the low temperature universality in glasses\cite{SS09}. It could also lead to an enhancement of our understanding of the microscopic nature of amorphous solids, and its relation to the detailed characteristization of the TLSs, and to the properties of amorphous solids at low temperatures.

It is exciting that the presence of weakly coupled TLSs leads to the dramatic reduction in the low energy DOS of the strongly coupled TLSs. In principle, this opens the opportunity to control and reduce the number of relatively strongly coupled TLSs at low energies by adding TLSs which are more weakly coupled, and in this way reduce the destructive absorption and decoherence effects caused by the TLSs. Although this suggestion does not seem to be easy practically because the nature of TLSs is unclear, one way of introducing weakly coupled TLSs can be associated with the hydrogenation of the material. Hydrogen atoms are expected to easily participate in tunneling and indeed they introduce the weakly coupled TLSs identified in aluminum and beryllium oxide glasses in Ref.\cite{KSG+13}. Moreover, hydrogenation of silicon oxide\cite{LWP+97} and silicon nitride\cite{PO10} results in the remarkable reduction of TLS induced absorption of sound or electromagnetic waves which can be due to the TLS gapping by hydrogen induced tunneling defects as described in this work. The investigation of this problem is a matter of current research.

Our results can also be carried through to magnetic insulators, as our model in Eqs.(\ref{H-St}),(\ref{Jcutoff}) describes the interactions of electronic and nuclear spins (the electron spin-spin interaction, the electron nuclear hyperfine interaction, and the nuclear spin-spin interaction are described by the first, second and third terms of Eq.~(\ref{H-St}), respectively). In particular, our results suggest a remarkable reduction in electronic spin flip rate in random magnetic systems at temperatures corresponding to the thermal energy being smaller than the typical hyperfine interaction. This implies a corresponding reduction in the decoherence of spin qubits at very low temperatures, and is a subject of a separate study.

All our considerations above assume that the dominant TLS-TLS interaction is acoustic rather than electric. While this may generally be the case, the smallness of the acoustic interactions between tau TLSs opens the possibility that electric dipolar interactions may be stronger than the acoustic interactions for these acoustically weakly interacting TLSs.
This possibility was recently discussed in Ref. \cite{CMBS21}. It was shown to lead, for the tau-TLSs, to a deeper dip in their DOS at low energies, accompanied by a power-law-like energy dependence at low energies, compatible with some experimental observations.

The datasets generated during and/or analysed during the current study are available from the corresponding author on reasonable request.

\section{Acknowledgments}

We would like to thank Ariel Amir, Juan Carlos Andresen, Danny Barash, Doron Cohen, Helmut Katzgraber, and the late Yoseph Imry, for useful discussions. M. S. acknowledges financial support from the ISF (Grant No.
2300/19). A. B. acknowledges the support by Carrol Lavin Bernick Foundation Research Grant (2020-2021), NSF CHE-2201027 grant and LINK Program of the NSF and Louisiana Board of Regents.

\appendix

\section{Calculation of the uncorrelated $\tau$ DOS}
\label{Sec:tauDOS}

\begin{figure*}[!tbh]
\centering
\subfigure[]{
  \includegraphics[width=8.6cm]{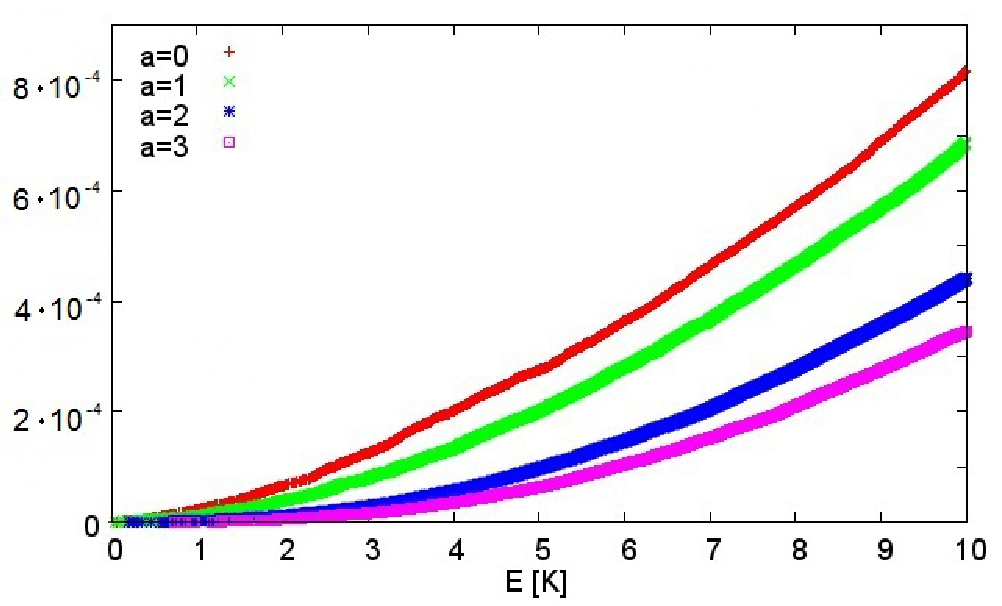}
   \label{fig:Fig3subfig1}
   }
 \subfigure[]{
  \includegraphics[width=8.6cm]{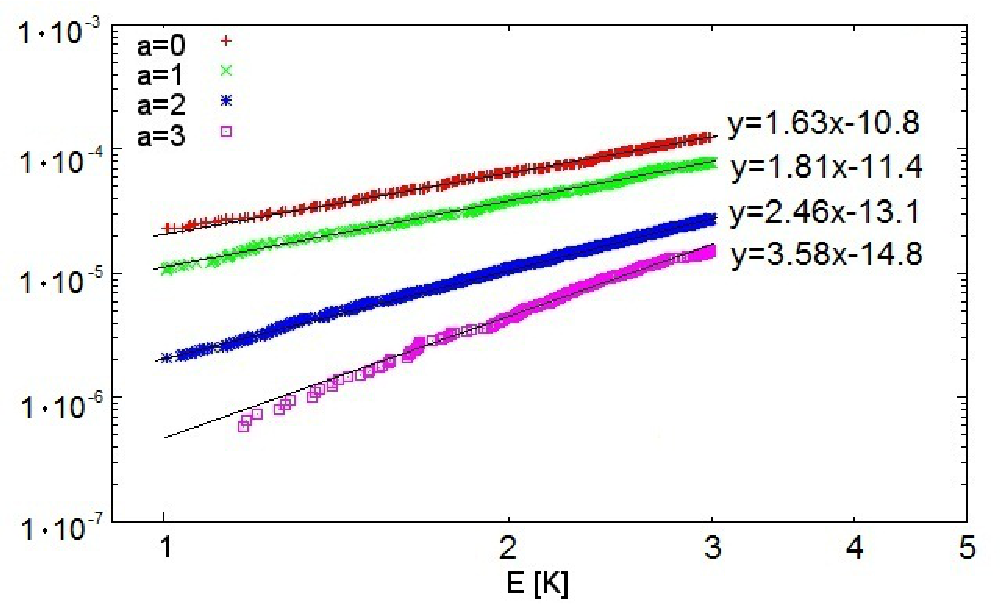}
   \label{fig:Fig3subfig2}
  }
\caption{ The integrated $S$-TLSs DOS for a cube of $14\times14\times14$ with $\rho=0.25$ and different values of interaction cutoff, $\tilde{a}=$0, 1, 2, 3. (a) Integrated $S$-TLSs DOS for $0 < E < 10$K.  (b) Log-log plot of the integrated $S$-TLSs DOS in low energy range along with best linear fits.
}
\label{fig:Fig3}
\end{figure*}

In this Appendix we calculate explicitly the variance, and discuss the functional form, of the distribution of the single particle excitation energies of the $\tau$-TLSs, neglecting the Efros-Shklovskii type correlations.

\begin{equation}
\langle E^2_{\tau} \rangle =\int_{-\infty}^{\infty} d\eta \frac{1}{\sqrt{2\pi}}e^{-\eta^2/2} \eta^2 \rho_S \sum_j  \frac{(2gJ_o)^2}{(R_j^3+\tilde{a}^3)^2}
\label{ntausum}
\end{equation}
where $j$ denotes all sites on the lattice except the origin. We consider here only the dominant $S-\tau$ interaction. The impurities are placed randomly on the lattice sites, with density $\rho_S$. However, for the averaging we can assume all sites are occupied and multiply by $\rho_S$.

Let us first consider the case $\tilde{a} >> a_o$ (and take $a_o=1$). In this limit we can approximate

\begin{equation}
\langle E^2_{\tau} \rangle = \rho_S \int_{-\infty}^{\infty} d\eta \frac{1}{\sqrt{2\pi}}e^{-\eta^2/2} \eta^2 \int_0^{\infty} 4\pi R^2dR \frac{(2gJ_o)^2}{(R^3+\tilde{a}^3)^2} \, .
\end{equation}
Performing the integral and taking the square root we obtain for the standard deviation of the distribution we obtain

\begin{equation}
E_{\tau}^{typ}=\sqrt{\frac{16\pi}{3} \frac{\rho_S}{\tilde{a}^3}} g J_o \, .
\label{Etyp}
\end{equation}

The calculation above allows for the $\tau$-TLSs to have either positive or negative energies. At zero temperature, however, all TLSs are at their ground states, and we are interested in the distribution of their excitation energies. This is given by the positive part of the Gaussian distribution, multiplied by two, as all negative values become positive (physically, a negative value means that a $\tau$-spin is placed in its high energy state, and thus needs to be flipped).
The peak value of the distribution at $E=0$ is then obtained

\begin{equation}
n_{\tau}^0(0)= \frac{2 \rho_{\tau}}{\sqrt{2\pi} E_{\tau}^{typ}} =  \frac{\sqrt{3} \sqrt{(\rho_{\tau}^2/\rho_S) \tilde{a}^3}}{4 \sqrt{2} \pi gJ_o} = \frac{\sqrt{3} \sqrt{\rho \tilde{a}^3}}{4 \sqrt{2} \pi gJ_o}
\label{DOStauzero}
\end{equation}
where in the last equation we assume for the spatial densities of the TLSs $\rho_S=\rho_{\tau}\equiv \rho$.

For the calculation of the uncorrelated distribution of the single particle excitations of the $S$-TLSs the same equations hold replacing $g$ with unity and $\rho_{\tau}$ with $\rho_S$.

If the condition $\tilde{a} \gg a_o$ is not fulfilled, one needs to evaluate the sum in Eq.(\ref{ntausum}) explicitly. For $\tilde{a}=0$ one finds $E_{\tau}^{typ}(\tilde{a}=0)=5.797 \sqrt{\rho} gJ_o$ and $n_{\tau}(0)=0.138 \sqrt{\rho}/(gJ_o)$.  Exact results for $E_{\tau}^{typ}$ for various cutoffs are given in Table \ref{tab:Table1}, and compared to the approximate values calculated with Eq.(\ref{Etyp}).

\vspace{0.2cm}

\subsection{Functional form of the uncorrelated $\tau$ DOS}

In this section we discuss the condition for the functional form of the $\tau$-TLS DOS to be nearly a Gaussian. For this we have to show that

\begin{equation}
\frac{\langle E_\tau^4 \rangle}{\langle E_\tau^2 \rangle^2} - 3 \ll 1 \, .
\end{equation}

The Kurtosis of the energy of a $\tau$-TLS can be written as

\begin{equation}
\langle E_\tau^4 \rangle = \sum_{ijkl} J_{oi} J_{oj} J_{ok} J_{ol} \eta_{oi} \eta_{oj} \eta_{ok} \eta_{ol} \rho_i \rho_j \rho_k \rho_l
\end{equation}
where $\eta_{oi} J_{oi}$ is the interaction of a $\tau$-TLS at site $o$ with an $S$-TLS at site $i$, $\eta_{oi}$ is a random Gaussian variable with unity variance, and

\begin{equation}
J_{oi} \equiv \frac{(2gJ_o)}{(R_i^3+\tilde{a}^3)} \, .
\end{equation}
$\rho_i$ is the occupation ($0$ or $1$) of site $i$, and $\langle \rho_i \rangle = \rho$.

Standard averaging over the random interaction and occupation variables gives

\begin{widetext}

\begin{equation}
\langle E_\tau^4 \rangle = \sum_{ijkl} J_{oi} J_{oj} J_{ok} J_{ol} \{[\delta_{ij}\delta_{kl}(1-\delta_{jk}) + \delta_{ik}\delta_{jl}(1-\delta_{jk}) + \delta_{il}\delta_{jk}(1-\delta_{jl})] \langle \eta^2 \rangle^2 \rho^2 + \delta_{ij} \delta_{jk} \delta_{kl} \langle \eta^4 \rangle \rho \}
\end{equation}
\end{widetext}
Since $\eta$ is a gaussian random variable, $\langle \eta^2 \rangle =1 $ and $\langle \eta^4 \rangle =3 $, resulting in

\begin{equation}
\langle E_\tau^4 \rangle = 3\rho^2 \sum_{ik} J_{oi}^2 J_{ok}^2 + 3\rho(1-\rho)\sum_i J_{oi}^4
\label{Eq:Kurtosis_f}
\end{equation}
The first term on the right hand side of Eq.(\ref{Eq:Kurtosis_f}) is three times the square of the variance. The deviation of the distribution from Gaussian can be therefore estimated by the ratio of the two terms of the Kurtosis. Straight forward integration shows this ratio to be $(1-\rho)/(4 \pi \rho \tilde{a}^3)$, and therefore negligible for $\rho \tilde{a}^3 \gtrsim 0.1$. This is indeed depicted in Fig.1(a). This condition is equivalent to the left side of the condition $[z(1+\tilde{a}^3)]^{-1} < \rho \lesssim 1$ which appears in the paragraph above Eq.(\ref{E-typ}) in the main text. The right side of this equation ($\rho \lesssim 1$) ensures the validity of the assumption of random interactions used in the present derivation. Moreover, the excellent agreement between our results here and the Monte Carlo results presented in Sec.\ref{Sec:Numerical} support the validity of the description of the $\tau$-TLSs via the effective Hamiltonian (\ref{tauRFIM}). Note, that if $\rho \tilde{a}^3 \lesssim 1$ our arguments above hold for the typical energy, but the distribution deviates from a Gaussian\cite{BH77}.

\begin{table}
\def\arraystretch{1}
\begin{tabular}{|c|c|c|c|}
	\hline
$\rho$ & $\tilde{a}$  &  $E_{\tau}^{typ}$ [exact] & $E_{\tau}^{typ}$ [Eq.(\ref{Etyp})]  \\
	\hline
0.5 &	0 &	4.099 $gJ_o$ & n.a. \\
0.5 &	1 &	2.474 $gJ_o$ &	2.894 $gJ_o$ \\
0.5 &	2 &	1.008 $gJ_o$ &	1.023 $gJ_o$ \\
0.5  & 	3 &	0.554 $gJ_o$ & 0.557 $gJ_o$ \\
	\hline
\end{tabular}
\caption { Standard deviation $E_{\tau}^{typ}$ of the distribution of $n_{\tau}(E)$ for various cutoffs. Results are given for $n=0.5$. Results for other spatial concentrations $n$ are easily deduced since both exact and approximate values are proportional to $\sqrt{\rho}$.}
\label{tab:Table1}
\end{table}

\section{Analytical calculation of the S-TLS DOS}
\label{Sec:SDOS}

In this appendix we derive the low energy density of states of the $S$-TLSs as is given in Eqs.(\ref{PE6}),(\ref{alphaanalytical}) in the main text, starting from Eq.(\ref{PE2}).

We start by performing the integration $d\zeta$. Using the fact that the $\Theta$ function reduces to the condition

\begin{equation}
\zeta<\frac{4\eta gJ_o}{E+E_{\tau}}-\tilde{a}^3
\end{equation}
we obtain

\begin{multline}
P(E)=exp\biggl[-\frac{4\pi}{3}\int_{\frac{E\tilde{a}^3}{4gJ_o}}^{\infty} d\eta \frac{1}{\sqrt{2\pi}} e^{-\eta^2/2}\\
\cdot \int_0^{4\eta gJ_o/\tilde{a}^3} dE_{\tau} n_{\tau}(0) \left(\frac{4\eta gJ_o}{E+E_{\tau}} - \tilde{a}^3\right)\biggr] \, .
\label{PE3}
\end{multline}
Defining $E`_{\tau}\equiv E_{\tau}+E$ and performing the integration $dE`_{\tau}$ we obtain, up to logarithmic accuracy,

\begin{multline}
P(E) \approx exp\biggl[-\frac{4\pi}{3} n_{\tau}(0) \int_{\frac{E\tilde{a}^3}{4gJ_o}}^{\infty} d\eta \frac{1}{\sqrt{2\pi}} e^{-\eta^2/2} \\
\cdot 4\eta gJ_o \log{\frac{4\eta gJ_o}{\tilde{a}^3 E}} \biggr] \, .
\label{PE4}
\end{multline}
For $E \ll 4g J_o/\tilde{a}^3$, replacing inside the log $\eta=1$ and performing the $\eta$ integration we obtain

\begin{equation}
P(E) \approx exp\left[-\frac{16\pi n_{\tau}(0) gJ_o}{3 \sqrt{2\pi}} \log{\frac{4 gJ_o}{\tilde{a}^3 E}} \right] \, .
\label{PE5}
\end{equation}
Using the definition in Eq.(\ref{PEdef}), the form of $n_S(E)$ in Eq.(\ref{PE6}) with the power $\alpha$ in Eq. (\ref{alphaanalytical}) are readily obtained.

\section{Numerical evaluation of the power $\alpha$}
\label{Appnumerical}

\begin{table}
\def\arraystretch{1}
\begin{tabular}{|c|c|c|c|c|}
	\hline
$\rho$ & $\tilde{a}$  &  L & $\alpha$ numerical  & $\alpha$ analytical   \\
	\hline
0.25 & 0 &	14 &	0.63$\pm$0.14 &  0.46\\
0.25 & 1 &	14 &	0.81$\pm$0.21  &  0.76\\
0.25 &	2 &	14 &	1.46$\pm$0.26 &  1.87\\
0.25 &	3 &	14 &	2.44$\pm$0.18 &  3.40 \\
0.5 &		0 &	12 &   0.78$\pm$0.04 & 0.65\\
0.5 &		1 &	12 &	1.05$\pm$0.1 & 1.08\\
0.5 &		2 &	12 &	2.09$\pm$0.13 & 2.65\\
0.5 &		3 &	14 &	3.98$\pm$0.69 & 4.81\\
	\hline
\end{tabular}
\caption {Values of the power $\alpha$ for the $S$-TLSs DOS for $\rho=0.25, 0.5$ and different values of interaction cutoff, $\tilde{a}=$0, 1, 2, 3. Numerical values are obtained from the best fits to the corresponding integrals plots. Analytical value are obtained using Eq.(\ref{alphaanalytical}). }
\label{tab:Table2}
\end{table}

\begin{figure}[t]
\centering
\includegraphics[width=8.6cm]{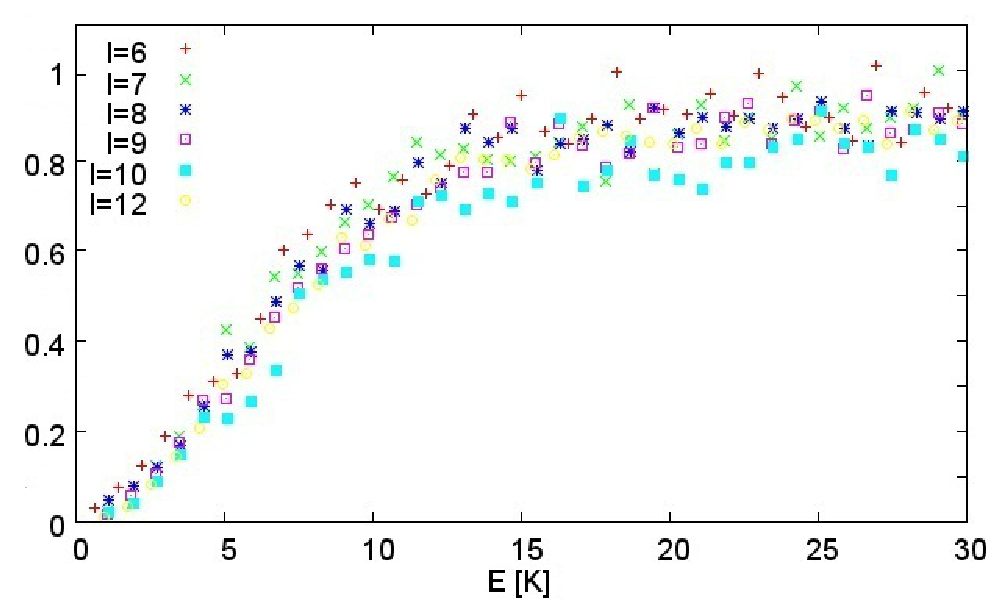}
\caption{ The ratio of $n_S(E)$ in the presence of $\tau$ TLSs to $n_S(E)$ in the absence of $\tau$ TLSs for L=6,7,8,9,10,12. $\tilde{a}=2$, $\rho=0.5$.}
\label{fig:sizes}
\end{figure}

In Sec. \ref{Sec:Analytical} we derived analytically the power $\alpha$ of the $S$-TLS DOS at low energies. Here we extract this power numerically by analyzing the integrated DOS for various values of the interaction cutoff and for $\rho=0.25$ and $\rho=0.5$.
The integrated DOS is plotted in Fig.\ref{fig:Fig3} for a cube of volume 14x14x14, with $\rho=0.25$ and $\tilde{a}=0,1,2,3$. The same data is plotted in a log-log scale in Fig.\ref{fig:Fig3}(b). The power $\alpha$ can in principle be obtained by a linear fit to the data at low energies, and then subtracting unity from the obtained slope. Because of finite size effects, and scarcity of the data at the very low energies, we approximate $\alpha$ by fitting the data at the energy range of $1-3$K. We determine the statistical errors using a bootstrap analysis. The same procedure is then repeated for $\rho=0.5$.
We present our results in Table \ref{tab:Table2},
and compare them to our analytic results.
Note that the numerical results are in reasonable agreement with the analytically obtained functional dependence of $\alpha \propto \sqrt{\rho}$, and for cutoffs $\tilde{a}=2,3$ with the functional dependence of $\alpha \propto \sqrt{\rho \tilde{a}^3}$ predicted analytically for $\tilde{a} \gg 1$.

\vspace{0.2cm}

\section{Effects of finite size}
\label{Appfinitesize}

Our numerical results deviate from the analytical results because of finite size effects.
These finite size effects can be directly seen in Fig.\ref{fig:sizes}, where we plot the ratio of $n_S(E)$ in the presence of $\tau$ TLSs to $n_S(E)$ in the absence of $\tau$ TLSs for L=6,7,8,9,10,12, short distance cutoff $\tilde{a}=2$, and concentration $n=0.5$. We clearly see that with increasing size the gap becomes deeper, with size dependence becoming less appreciable at the larger sizes.


\begin{thebibliography}{10}
\bibitem{Lorenote}
The distribution is Gaussian for the regime of intermediate defect concentrations which is our main interest here. For the extreme dilute regime a similar effective random field model would render a Lorentzian distribution, as rare near neighbors will dictate the distribution at large energies. This change of distribution, however, would affect our results for the $S$-TLSs only quantitatively.
\end{thebibliography}
\end{document}